\documentclass[aps,prl,superscriptaddress,twocolumn,floatfix,showpacs]{revtex4}
\usepackage{amsfonts}
\usepackage{color}
\usepackage{graphicx}
\usepackage{subfigure} 
\usepackage{dsfont}
\newcommand{\beq}{\begin{equation}}
\newcommand{\eeq}{\end{equation}}
\newcommand{\beqa}{\begin{eqnarray}}
\newcommand{\eeqa}{\end{eqnarray}}
\usepackage{graphicx,amssymb,amsmath,subfigure}
\usepackage{dcolumn}
\usepackage{bm}

\begin{document}
\title{An Example of Time Reversal Invariant Kerr Effect }

\author{Alberto Cortijo}
\email{alberto.cortijo@csic.es}
\affiliation{Instituto de Ciencia de Materiales de Madrid,
CSIC, Cantoblanco; 28049 Madrid, Spain.}
\begin{abstract}
Here we describe how certain classes of two dimensional topological insulators, including the CdTe$/$HgTe quantum wells, display a new type of optical activity in two dimensions similar to the magneto-optical Kerr effect in the quantum Hall effect. This optical activity is characterized by a genuine Kerr angle and it is compatible with time reversal symmetry, being thus fundamentally different to other known types of time reversal invariant optical activity. The term responsible of such optical activity, having the form of $(\mathbf{E}\cdot\partial\mathbf{B}/\partial t-\mathbf{B}\cdot\partial\mathbf{E}/\partial t)$, can be considered a time reversal invariant counterpart of the magneto-electric term $\mathbf{E}\cdot\mathbf{B}$. The microscopical origin of this response is a chiral non-minimal coupling between electrons and the external electromagnetic field. This optical activity constitutes a proof of principle that there is possible to find systems that are time reversal invariant displaying a genuine Kerr effect.
\end{abstract}

\maketitle

Nowadays it is widely accepted that topological insulators in two dimensions (2DTI) like the Quantum Spin Hall effect (QSHE) are the time reversal invariant counterparts of the quantum Hall effect (QHE)\cite{HK10,QZ11}. Like the latter, 2DTIs show up some properties that distinguish them from ordinary insulators, like pairs of counter-propagating, symmetry protected one dimensional edge states\cite{KM052,BZ06}. The appearance and robustness of such states stems from the non trivial topological structure displayed by the electronic band structure, as in the case of the QHE\cite{KM05}. 

The QHE shows its non trivial topological properties in other ways than the presence of one dimensional edge states. Specifically it exhibits an specific form of \emph{optical activity} in its electronic response to external electromagnetic fields: The magneto-optic Kerr and Faraday effects (for an authoritative survey into the subject, we refer to \cite{LLP60}.): When a two dimensional electron gas (2DEG) is placed in an external magnetic field $\mathbf{B}$ the electromagnetic response of the system acquires a term that is antisymmetric and proportional to a \emph{gyrotropic} vector $g_{l}$, $D_{i}=\epsilon^{0}\delta_{ij}E_{j}+\varepsilon_{ilj}g_{l}E_{j}$. The gyrotropic vector $g_{l}$ appears to be proportional to the external magnetic field. Also, alternatively, this response can be written in the form of a magneto-electric term in the constitutive relations\cite{HC84,QHZ08}. 

From the point of view of the electronic structure, the reason of the existence of this term is the appearance of a transverse component in the electric current of the 2DEG, $ J^{i}_{e}=\sigma_{xy}\epsilon_{ij}E_{j}$, where $\sigma_{xy}$ is the Hall conductivity	\cite{L81}. It is by now widely understood that the constraint of invariance under time inversion in 2DTI implies a zero net transverse electric current $ J^{i}_{e}$ but a non-vanishing spin (or pseudo-spin) current: $ J^{i}_{s}\equiv J^{i}_{\uparrow}-J^{i}_{\downarrow}=2\sigma_{xy}\epsilon_{ij}E_{j}$ appearing in these systems (the subscripts $\uparrow,\downarrow$ label the two species of electrons in the 2DTI, like the spin or the third component of the total angular momentum)\cite{KM05,BHZ06}. However, despite of the relevance of the existence of a non-vanishing spin current in 2DTIs, it seems obvious that the current entering in the Maxwell equations is not $ J^{i}_{s}$ but $ J^{i}_{e}$ which is zero. In view of the preceding discussion it might appear that 2DTIs do not show any optical activity similar to the Kerr effect in their electromagnetic response.

Actually, while all 2DTI share the general property of having a non-zero $J^{i}_{s}$, the condition $J^{i}_{e}=0$ might not need to be necessary although time reversal symmetry could seem to demand it. The reason why it does not need to be the case is that electrons in a 2D system can respond to external electromagnetic fields not only through the usual Lorentz force (coming from the minimal coupling between electrons and the electromagnetic gauge field) but also by other external forces thus inducing electric current in the system. Examples of this are the Thermal Hall effect and valley Hall effect\cite{LP81,GKG10,VAA13}. However the observability of these induced currents are subject to the presence of the associated driving force, making the response sensitive to other degrees of freedom that could blurry the pure electromagnetic response.

Even so, recently it has been pointed out that another force related to external electromagnetic field and different to the Lorentz force can act on the electrons in CdTe/HgTe quantum wells (QW)\cite{C14}. This force stems from a non minimal coupling between the local dipole moments of the system and the electric field. This driving force will induce a net electronic current and it is proportional to the externally applied electric field, thus avoiding the problems mentioned above. 

Microscopically, the origin of this effect comes from the peculiar low energy band structure around the $\Gamma$ point of a CdTe$/$HgTe QWs in the so-called inverted regime\cite{BHZ06,KWS07}. The CdTe$/$HgTe QW sub-band structure is made by the states $|E1,m_{J}=\pm\frac{1}{2}\rangle=\hat{\alpha}|\Gamma_{6},\pm\frac{1}{2}\rangle+\hat{\beta}|\Gamma_{8},\pm\frac{1}{2}\rangle$ and $|H1,m_{J}=\pm\frac{3}{2}\rangle=|\Gamma_{8},\pm\frac{3}{2}\rangle$\cite{MR11}. The presence of both $\pm$ signs is required by time reversal invariance. Importantly, the state $|\Gamma_{6},\pm\frac{1}{2}\rangle$ is a s-like state ($l=0$) while the state $|\Gamma_{8},\pm\frac{3}{2}\rangle$ is a p-like state with orbital angular momentum $l=\pm1$ (the degeneracy between the levels $l=\pm1$ is lifted by the large value of the spin orbit coupling in these systems). 

While the states $|H,\pm\rangle$ and $|E,\pm\rangle$ are confined states within the QW structure, they are Bloch states along the plane defining the QW, implying that they can be conveniently written as linear combinations of the atomic p-like and s-like wave functions respectively within a tight-binding scheme. Then the important observation is that the hopping terms between $|H,\pm\rangle$ and $|E,\pm\rangle$ of neighbouring sites must have a vector character $l=\pm1$ to preserve angular momentum in the hopping process, while hopping between the same type of states between neighbours have scalar ($l=0$) character. This is the reason why, when the on-site atomic energies are taken into account, the low energy $\mathbf{k}\cdot\mathbf{p}$ Hamiltonian has the form of a massive Dirac Hamiltonian\cite{BHZ06}.

When an external (time dependent) electric field $\mathbf{E}(t)$ is applied along the QW plane, the electrons experience the standard Lorentz force in the form of the usual quantum mechanical coupling (Peierls coupling) between the electronic states and the gauge vector field, $\mathbf{k}\rightarrow\mathbf{k}-e\mathbf{A}$. However, now because the electric field is an operator with vector character, it can induce local electric dipole transitions between the s-like and p-like components of $|E,\pm\rangle$ and $|H,\pm\rangle$ respectively, thus appearing an extra coupling between the electromagnetic field and the on-site electric dipole moment of the form $e\langle s,\pm\frac{1}{2}|\mathbf{x}|p,\pm\frac{3}{2}\rangle\cdot \mathbf{E}$, according to the standard selection rules for electric dipole transitions. Here, time reversal symmetry requires that $\langle s,+\frac{1}{2}|\mathbf{x}|p,+\frac{3}{2}\rangle=-\langle s,-\frac{1}{2}|\mathbf{x}|p,-\frac{3}{2}\rangle$, that is, the states with opposite $m_{J}$ projection that are related by time reversal symmetry, couple with opposite sign to the external electric field through the dipole term. This observation is the key for the rest of the discussion.

In a tight-binding scheme and within the dipole approximation the external electric field will couple to the electrons through a term of the form $H_{int}=-e\int d^{2}\mathbf{r}\mathbf{E}(t)\cdot\mathbf{r}$, that, after going to the tight-binding basis, translates into $H_{i}=-e\sum_{i,\alpha,\beta}\mathbf{E}(t)(\delta_{\alpha\beta}\mathbf{R}_{i}+\mathbf{d}_{\alpha\beta})c^{*}_{\alpha,i}c_{\beta,i}$. The first term can be written as a Peierls phase in the electronic Hamiltonian in the usual way (after making a gauge transformation). However there is still the term $\mathbf{d}_{\alpha\beta}\cdot\mathbf{E}(t)$ that represents a non-minimal local Stark term that directly couples the electrons to the electric field, that is, the complete electron-electromagnetic field coupling is, in the dipole approximation ($\mathbf{A}(t,\mathbf{q})\sim \mathbf{A}(t)\delta_{\mathbf{q,0}}$):
\begin{eqnarray}
H_{int}= &-&e\sum_{\mathbf{k},\alpha\beta}\frac{\partial h_{\alpha\beta}(\mathbf{k})}{\partial \mathbf{k}}\cdot\mathbf{A}(t)c^{*}_{\alpha,\mathbf{k}}c_{\beta,\mathbf{k}}-\nonumber\\
&-&e\sum_{\mathbf{k},\alpha\beta}\mathbf{E}(t)\cdot\mathbf{d}_{\alpha\beta}c^{*}_{\alpha,\mathbf{k}}c_{\beta,\mathbf{k}},\label{totalHam}
\end{eqnarray}
where $h_{\alpha\beta}$ represents the matrix elements of the Hamiltonian in momentum space. The second coupling term in equation (\ref{totalHam}) is a type of non-minimal coupling term in Condensed Matter Physics. Other common examples of non-minimal terms are, for instance, the Zeeman coupling between an external magnetic field $B_{i}$ and the electron spin: $H_{Z}=\frac{\mu_{B}g}{2}B_{i}c^{+}_{s}\sigma^{i}_{ss'}c_{s'}$. The low energy effective theory around the $\Gamma$ point takes the well known form of the two dimensional massive Dirac theory\cite{BHZ06} coupled to the electromagnetic field $A_{i}$ in the standard fashion, $ev\bar{\psi}\gamma^{i}\psi A_{i}$ ($v$ is the Fermi velocity of the electrons near the $\Gamma$ point), and coupled also to a chiral field denoted by $V_{i}$, $ev\bar{\psi}\gamma^{i}\gamma_{S}\psi V_{i}$ ($\gamma_{S}$ is a chirality matrix, see the Supplementary Information for details), that turns out to be proportional to the electric field: $V_{i}=\zeta^{0}E_{i}$, where $\zeta^{0}=d/v$, with $d=\langle s,+\frac{1}{2}|x|p,+\frac{3}{2}\rangle$. It is not easy to give a precise value of the parameter $d$. However, within the tight-binding scheme, we can estimate its value by the use of atomistic orbital wavefunctions. It turns out that a lower bound for the value of $d$ is of the order of the Bohr radius, $d\sim a_{0}\sim 0.5 \AA$. Knowing that the Fermi velocity $v$ in the inverted regime of CdTe/HgTe QW is $v\simeq 3.42$ eV$\AA$\cite{BHZ06}, the value of $\zeta^{0}$ is of the order of $\zeta^{0}\sim 0.14$ eV$^{-1}$ or larger. 

Let us study the electromagnetic macroscopic response derived from equation (\ref{totalHam}) by calculating the electronic current induced by an external electric field. If we set the chemical potential to lie within the electronic gap, the electromagnetic current appearing in the system in the long wavelength limit is, by direct computation \cite{C14},(See the supplementary material for details of the calculation.):
\begin{equation}
J^{i}_{e}=\frac{\alpha}{\pi}\sigma_{xx}(\omega)\delta^{ij}E_{j}-i\frac{\alpha}{\pi}\zeta^{0}\omega\sigma_{xy}(\omega)\varepsilon^{ij}E_{j}.\label{current}
\end{equation}


Let us stress that the current written in equation (\ref{current}) is defined \emph{within} the plane where the QW is, and it is induced only by the in-plane components of the electric field $E_{i}$ and its time derivative. The first term in equation (\ref{current}) corresponds to a more standard longitudinal current induced in the same direction that the external electric field. Also, the conductivities $\sigma_{xx}(\omega)$ and $\sigma_{xy}(\omega)$ are well known functions in the context of the physics of two dimensional massive Dirac systems\cite{R84,R842,TM10}. We have made explicit the appearance of the finite structure constant $\alpha$ in the above expressions for convenience.

Let us focus on the term proportional to $\sigma_{xy}$. From a low energy theory perspective, this term comes from the fact that the non-minimal coupling between the electronic dipole moment and the electric field is chiral, meaning that the two fermion species present in the system couple to the field $V_{i}$ with opposite sign\cite{C14}. Then the extra piece of the current can be derived from an effective Lagrangian that resembles the Chern-Simons Lagrangian but it is essentially different: $J^{i}_{e}=\delta S_{eff}/\delta A_{i}$, with 
\begin{equation}
S_{eff}=\int d^{3}xdt \delta(z)\left(\epsilon^{ilj}A_{i}\partial_{l}V_{j}+\epsilon^{ilj}V_{i}\partial_{l}A_{j}\right)-J^{i}_{e}A_{i},\label{effaction1}
\end{equation}
where the latin indices are for $(t,x,y)$. It is easy to see that the action defined in equation (\ref{effaction1}) is equivalent to the following one, written in terms of the electric and magnetic fields:
\begin{eqnarray}
S'_{eff}=-\int d^{3}xdt f(z) \sigma_{xy}\zeta^{0}\left( \mathbf{B}\cdot\frac{\partial\mathbf{E}}{\partial t}-\mathbf{E}\cdot\frac{\partial\mathbf{B}}{\partial t}\right),\label{effaction3}
\end{eqnarray}
up to a total divergence term that does not contribute to the dynamics. In obtaining equation (\ref{effaction3}) we have made use of the definition of the dual electromagnetic tensor $F^{*\mu\nu}=\epsilon^{\mu\nu\rho\sigma}\partial_{\rho}A_{\sigma}$ and the definition of the chiral field $V_{i}$. The function $f$ in equation (\ref{effaction3}) is a function satisfying $\partial_{\mu}f(z)=\delta(z)\delta_{\mu z}$.

Thus the macroscopic electric and magnetic dipole moments acquire an extra term of the form
\begin{subequations}
\begin{equation}
4\pi\Delta \mathbf{P}=-\sigma_{xy}f(z)\zeta^{0} \frac{\partial \mathbf{B}}{\partial t},
\end{equation}
\begin{equation}
4\pi \Delta \mathbf{M}=\sigma_{xy}f(z)\zeta^{0}\frac{\partial \mathbf{E}}{\partial t}.
\end{equation}\label{consteq}
\end{subequations}
We can make some comments about the form of the action in equation (\ref{effaction3}) and the constitutive relations defined in equation (\ref{consteq}). The action in equation (\ref{effaction3}) can be considered a time reversal invariant counterpart of the topological magneto-electric term $\theta\mathbf{E}\cdot\mathbf{B}$ appearing in three dimensional topological insulators\cite{QHZ08}. It is important to stress that the constitutive relations appearing in equation (\ref{consteq}) satisfy the Onsager reciprocity relations since, under time reversal inversion, $\sigma_{xy}\rightarrow-\sigma_{xy}$, $\zeta^{0}\rightarrow-\zeta^{0}$, while $\dot{\mathbf{B}}\rightarrow\dot{\mathbf{B}}$, and $\dot{\mathbf{E}}\rightarrow-\dot{\mathbf{E}}$ (remember that under time reversal inversion, $\mathbf{E}\rightarrow\mathbf{E}$ and $\mathbf{B}\rightarrow-\mathbf{B}$ and dotted quantities mean their time derivatives). The price to pay for being time reversal invariant is that, contrary to what happens with the topological magneto-electric term,  the coefficient $\sigma_{xy}\zeta^{0}$ is not quantized since $\zeta^{0}$ is a constitutive parameter that depends on the microscopic details of the system at hands.

We shall explore the experimental consequences of the extra terms of the constitutive equations by computing the transmitted and reflected electromagnetic waves by a QW\cite{FV09}. Let us consider an electromagnetic plane wave which travels through the QW sample placed in the $z=0$ plane. We will consider normal incidence for simplicity. 

To see how the new terms described in equations (\ref{consteq}) affect the optical activity in the system, we will restrict ourselves to frequencies smaller than the electron bandgap $\omega<2m$. In this frequency regime $Re[\sigma_{xx}]=0$, $Im[\sigma_{xy}]=0$ meaning that there is no dissipative component in the conductivities $\sigma_{xx}$ and $\sigma_{xy}$ due to production of electron-hole pairs. Also, it is known that the amplitudes for the transmitted ($t_{xx}, t_{xy}$) and reflected ($r_{xx},r_{xy}$) electromagnetic waves are well approximated by their leading values in an expansion in powers of the fine structure constant $\alpha$\cite{FV09}, making the discussion of the results easier. In order to study a Kerr-like effect we will put our attention to the behaviour of the reflected wave. The leading terms (in an expansion in small $\alpha$) of the polarization of the reflected wave are
\begin{subequations}
\begin{equation}
r_{xx}\simeq-2i\frac{\alpha}{\sqrt{\epsilon}} Im[\sigma_{xx}], 
\end{equation}
\begin{equation}
r_{xy}\simeq -2i\frac{\alpha}{\sqrt{\epsilon}} \zeta^{0}\omega Re[\sigma_{xy}].
\end{equation}\label{refcoeff}
\end{subequations}
For a thickness of $7$nm the system is in the inverted (topologically non trivial) regime with a bandgap is $2m=13.7$ meV\cite{KWS07}, corresponding to frequencies in the range of THz. Also, electromagnetic waves in this range posses a wavelength of the order of hundreds of microns, much larger than the thickness of the QW (of the order of few nm\cite{KWS07}) justifying the approximation of the QW as an ideal plane. In equations (\ref{refcoeff}) the values of the reflection coefficients are sensitive to the value of the dielectric constant $\epsilon$ of the medium surrounding the QW. In our case, this medium is assumed to be three dimensional CdTe with a dielectric constant\cite{GHF58} $\epsilon\sim10$. 

\begin{figure}[t]
 \centering
(a)
\includegraphics[width=0.5\textwidth]{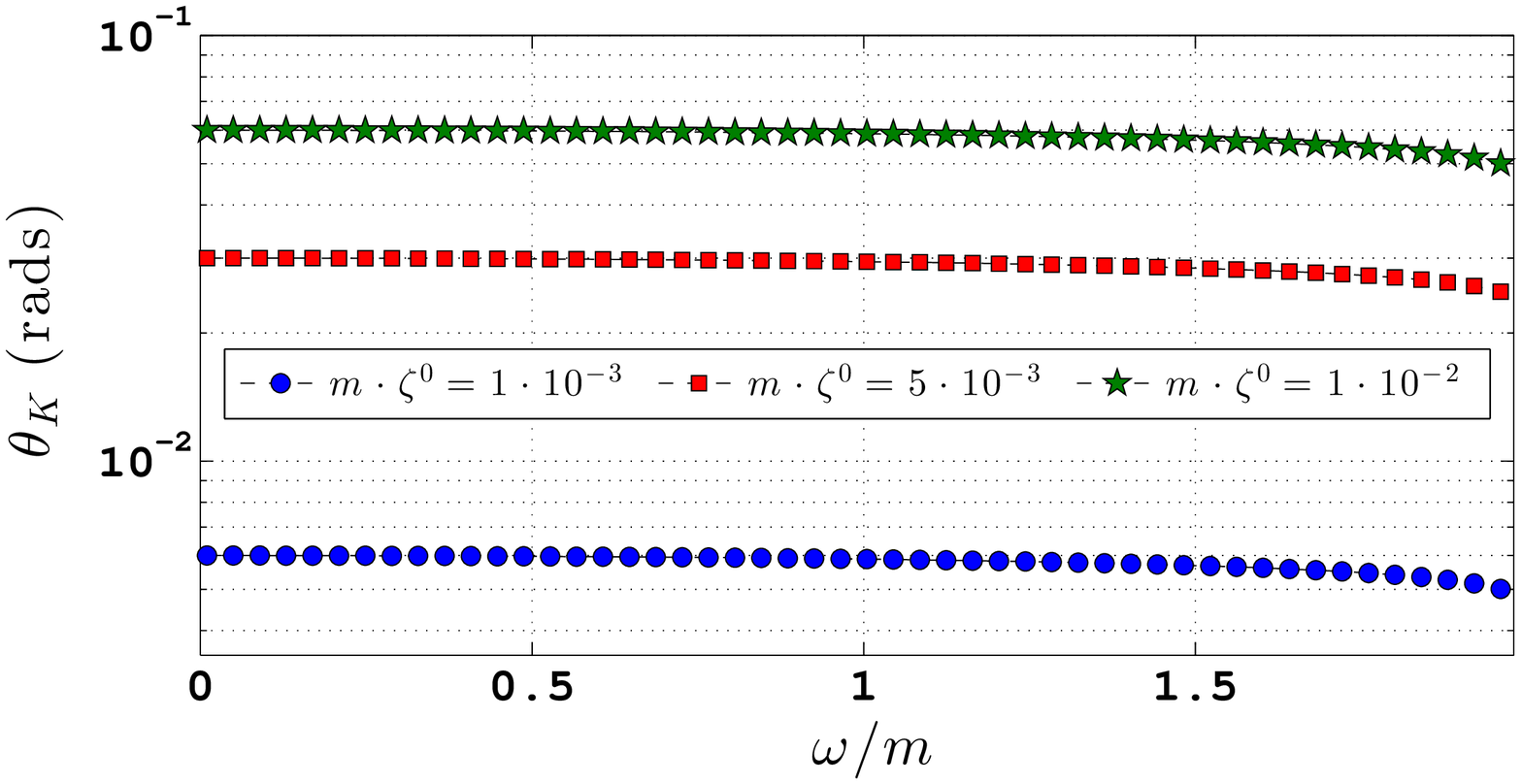}
(b)
\includegraphics[width=0.5\textwidth]{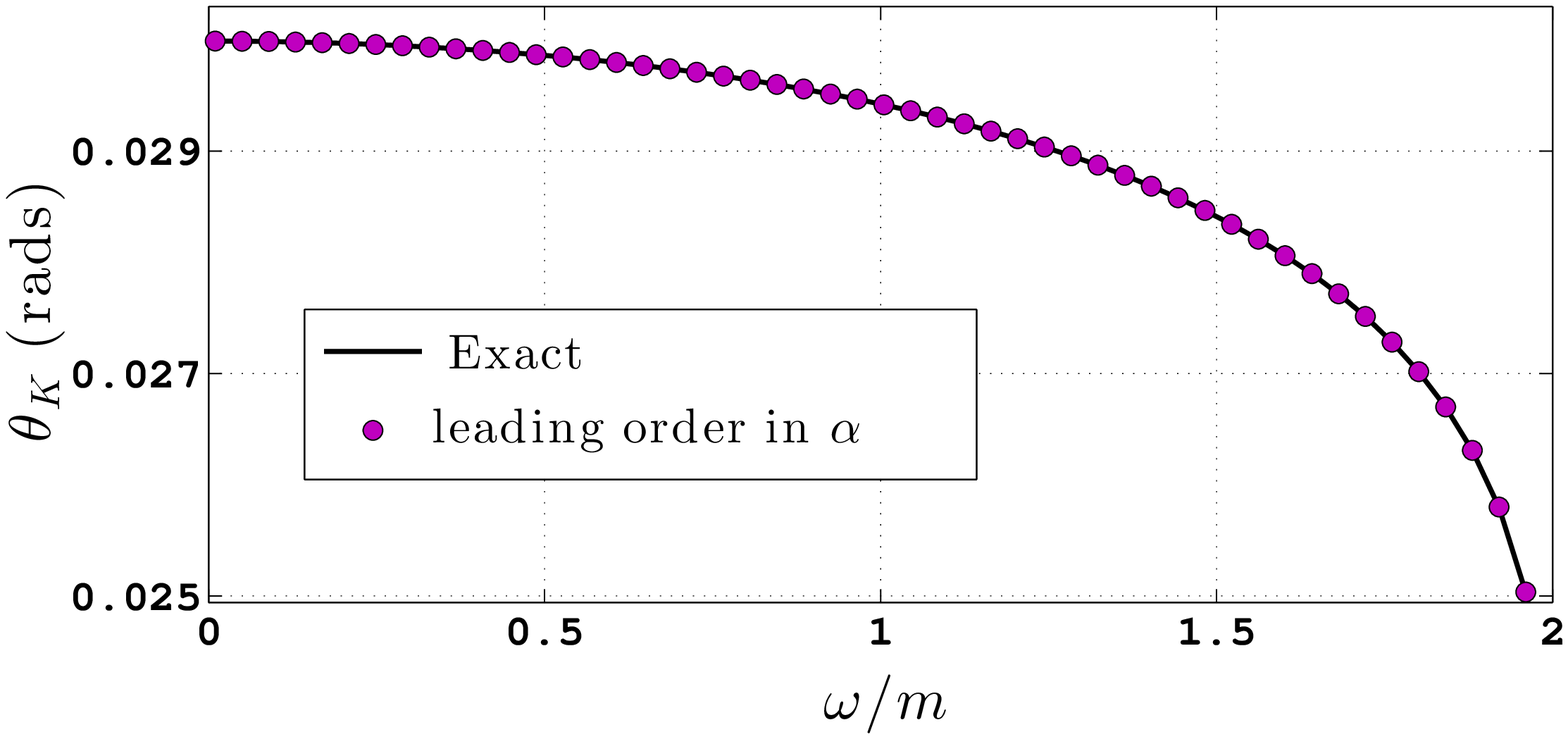}
\caption{(Colour online) (a) Behaviour of the Kerr angle $\theta_{K}$ as a function of the normalized frequency $\omega/m$ for different values of the dimensionless parameter $m\zeta^{0}$. The vertical axis is presented in logarithmic scale.The Kerr angle remains almost constant for a wide range of frequencies below the electron-hole pair production threshold. (b) Comparison between the exact expression for $\theta_{K}$ and the leading order given by equation (\ref{approxthetak}) for the value $m\cdot\zeta^{0}=5\cdot 10^{-3}$.}
\label{fig:combined}
\end{figure}


The first observation is that both components of the polarization of the reflected wave are complex. It means that the reflected wave is in antiphase with respect to the incident wave, that is, if the oscillating part of the incident wave is, say, $\cos(\omega t-q z)$, the reflected wave will be quite accurately of the form $\sin(\omega t-q z)$. Also, being both coefficients imaginary to leading order in $\alpha$, the polarization will be approximately linear instead of elliptical. Then, the rotation of the polarization of the reflected wave is
\begin{equation}
\theta_{r}\simeq\arctan\left(\zeta^{0}\omega\frac{Re[\sigma_{xy}(\omega)]}{Im[\sigma_{xx}(\omega)]}\right).\label{approxthetak}
\end{equation}
It is clear that when $\zeta^{0}$ is zero, there is no rotation of the polarization plane of the electromagnetic wave, as should be. In Fig.(\ref{fig:combined},a) we have plotted $\theta_{r}\equiv\theta_{K}$ as a function of the normalized frequency $\omega/m$ of the incident wave for frequencies below the electron-hole pair production threshold. The value of $\theta_{K}$ ranges from values of the order of mrads to fractions of rad for changes of $m\cdot\zeta^{0}$ of one order of magnitude. These values are perfectly observable in current standard Kerr measurements\cite{SYY13}. In Fig.(\ref{fig:combined},b) it is compared the exact computed value of $\theta_{K}$ and the leading order term in $\alpha$. For frequencies smaller than the electron-hole pair production threshold, both results are in excellent agreement.

Recently, in basis of a previous discussion due to Halperin\cite{H92}, it has been argued that under the conditions of time reversal symmetry, equilibrium, and linear response there is no Kerr rotation\cite{HKK14,F14,PKY14}, so the result presented above might seem to be paradoxical, at least, since it is clear from the introductory discussion that the system we are using is time reversal invariant by construction, and the current in equation (\ref{current}) is computed within the linear response regime, implying that the quantities entering in equation (\ref{current}) are quantities defined in equilibrium. At this point it is instructive to make use of the following Maxwell equation $\nabla\times\mathbf{E}=-\dot{\mathbf{B}}$ ($c=1$), and rewrite the equation (\ref{consteq}a) for $\Delta\mathbf{P}$ in the form of 
\begin{equation}
4\pi\Delta \mathbf{P}=\sigma_{xy}f(z)\zeta^{0}\nabla\times\mathbf{E},\label{consteq3} 
\end{equation}
Written is this way, it might appear that the optical activity is similar to the optical activity that has been recently attributed in the literature to systems like cuprate superconductors. Precisely, nowadays it is becoming clear that the optical activity displayed by these systems does not lead to Kerr rotation of the polarization plane of the reflected wave. 

There are two important differences between the term of equation (\ref{consteq3}) and the actual expression of the polarization field $\mathbf{P}$ for these systems. For them, the constitutive relations are $4\pi \mathbf{P}=\gamma(z)\nabla\times\mathbf{E}+(1/2)(\nabla\gamma(z))\times\mathbf{E}$ and $\mathbf{H}=\mathbf{B}$. The presence of the two terms in this $\mathbf{P}$ results in the absence of Kerr rotation\cite{MY14}. In our case, the non trivial response leading to the equation (\ref{consteq}) is purely two dimensional and it appears only within the plane of the QW. In addition, we also have a term relating the magnetization $\mathbf{M}$ with the time derivative of $\mathbf{E}$ which is absent in the constitutive relations attributed to cuprate superconductors. 

Besides the difference between the constitutive relations, the absence of Kerr effect in time reversal invariant media is discussed in terms of the Lorentz reciprocity theorem. The statement of the absence of Kerr rotation under time reversal symmetry, linear response and equilibrium is based on the following symmetry of the electromagnetic propagator: $G^{A}_{\mu\nu}(\omega,\mathbf{r}_{2},\mathbf{r}_{1})=G^{A}_{\nu\mu}(\omega,\mathbf{r}_{1},\mathbf{r}_{2})$\cite{F14}. This symmetry property of the electromagnetic Green's function is tantamount to the elder formulation of the Lorentz reciprocity theorem\cite{LLP60,P04}:
\begin{equation}
\int d^{3}\mathbf{r}\left(\mathbf{E}_{1}\cdot\mathbf{J}_{2}-\mathbf{E}_{2}\cdot\mathbf{J}_{1}\right)=0,\label{reciprocity}
\end{equation}
so the statement is that for \emph{reciprocal} systems under the mentioned requirements and under the definition of reciprocity represented by the theorem (\ref{reciprocity}), there is no Kerr rotation. However, our system, despite of being time reversal invariant it is \emph{not reciprocal} y the previous sense. It is easy to show (see the last section of the supplementary material for the algebraic details of the derivation) that, under the same assumptions of external currents with harmonic time dependence the Lorentz reciprocity theorem (\ref{reciprocity}) is supplemented by an extra term:
\begin{eqnarray}
& &\int d^{3}\mathbf{r}\left(\mathbf{E}_{1}\cdot\mathbf{J}_{2}-\mathbf{E}_{2}\cdot\mathbf{J}_{1}\right)=\nonumber\\
&=&-2\omega^{2}\zeta^{0}\sigma_{xy}\int d^{3}\mathbf{r}f(z)\left(\mathbf{E}_{1}\cdot\mathbf{H}_{2}-\mathbf{E}_{2}\cdot\mathbf{H}_{1}\right).
\end{eqnarray}
Again, the physical origin of the non-reciprocity in the system described here is the chiral coupling between the electric field and the local dipole moments in the QW. Let us compare this result with the one we would encounter if we considered the Quantum Anomalous Hall effect (QAHE), which is the time reversal symmetry breaking counter-term of our system. The low energy electronic structure in the QAHE consists in a single specie of two dimensional massive Dirac fermions minimally coupled to the electromagnetic field. The presence of a single specie of Dirac fermions implies necessarily that time reversal symmetry is broken. Also, because there is precisely a single specie of Dirac fermions, the electromagnetic response is chiral leading to a Hall-like (non-reciprocal) optical activity. Now we turn to our system where the number of species of massive Dirac fermions is two and both have opposite chirality, as time reversal symmetry dictates. However these Dirac fermions couple to the external electromagnetic field also through an extra non-minimal coupling in a chiral fashion, that is, both species couple to the electric field equally. It means that, while for the standard minimal coupling the putative chiral effects coming from each fermion specie mutually cancel, this cancellation does not occur with this chiral non-minimal coupling, leading to a non-reciprocal activity fully compatible with time reversal symmetry.

Summarizing, we have defined a new type of optical activity in a time reversal invariant system like the CdTe/HgTe QW that has the same characteristics that the Kerr rotation that is perfectly measurable. The physical mechanism for such response is a chiral, non-minimal coupling between the intra atomic dipole moments in the QW and the external electric field. This coupling makes the response \emph{non-reciprocal}, avoiding the statement that the Kerr rotation is absent in time reversal invariant, linear systems in equilibrium.

The author gratefully acknowledges M. Sturla, M. A. H. Vozmediano, R. Aguado, and P San-Jose for illuminating discussions. The author aknowledges the JAE-doc program co–funded by the European Social Fund and the Ministerio de Economia y Competivididad and the Spanish MECD through Grants No. FIS2011-23713 and No. PIB2010BZ-00512 for financial support.


\clearpage
\begin{widetext}
\section{Supplementary information of ``An example of time reversal invariant Kerr effect"}

\subsection{Computation of the induced electronic current}
Here we will compute the induced electronic current in linear response, that is, to first order in all the external potentials. The low energy Hamiltonian of the electronic system in the QW, disregarding quadratic terms in momentum, has the form of the massive Dirac Hamiltonian\cite{BHZ06}:

\begin{equation}
H_{0}(\mathbf{k})=vs_{3}\tau_{1}k_{1}-vs_{0}\tau_{2}k_{2}+s_{0}\tau_{3}m.\label{DiracHam}
\end{equation}

Defining the adjoint spinor $\bar{\psi}=\psi^{+}\gamma^{0}$, with $\gamma^{0}=s_{0}\tau_{3}$, we can go into a Lagrangian description and define the following action, after coupling the electronic degrees of freedom to the electromagnetic gauge field and the electric field\cite{C14} (we will set $\hbar=1$ in the intermediate computations and restore it at the end):

\begin{equation}
\mathcal{S}=\mathcal{S}_{0}+S_{int}=\int d^{3}x \bar{\psi}\left(-i\gamma^{\mu}M^{\nu}_{\mu}\partial_{\nu}-m\right)\psi-e\bar{\psi}\gamma^{\mu}\psi M^{\nu}_{\mu}A_{\nu}(x)-e\zeta^{0}\bar{\psi}\gamma^{\mu}\gamma_{S}\psi M^{\nu}_{\mu}F_{0\nu}(x),
\end{equation}
where $F_{\mu\nu}=\partial_{\mu}A_{\nu}-\partial_{\nu}A_{\mu}$. The set of $\gamma$ matrices are $\gamma^{1}=is_{3}\tau_{2}$, $\gamma^{2}=is_{0}\tau_{1}$, and $\gamma_{S}=s_{3}\tau_{0}$, that defines the chirality of the two fermion species. In order to take into account the value of the Fermi velocity, we have defined the matrix $M^{\nu}_{\mu}=$diag$(1,v,v)$.

In perturbation theory, the induced electronic current reads

\begin{eqnarray}
J^{\mu}(x)&=&e\langle\bar{\psi}(x)\gamma^{\rho}\psi(x)\rangle M^{\mu}_{\rho}=\nonumber\\
&=&-eM^{\mu}_{\rho}\frac{1}{\mathcal{Z}_{0}}\int D\bar{\psi}\psi e^{iS_{0}}\bar{\psi}(x)\gamma^{\rho}\psi(x)[ 1+i\int d^{3}y e\bar{\psi}(y)\gamma^{\sigma}\psi(y)M^{\nu}_{\sigma}A_{\nu}(y)+\nonumber\\
&+&i\int d^{3}y e\zeta^{0}\bar{\psi}(y)\gamma^{\sigma}\gamma_{S}\psi(y)M^{\nu}_{\sigma}F_{0\nu}(y)+...].
\end{eqnarray}

Applying Wick's theorem, $J^{\mu}$ takes the following form in terms of the fermionic Green functions:

\begin{eqnarray}
J^{\mu}(x)&=&-i e^2\int d^{3}y\langle \gamma^{\rho} G_{0}(x-y)\gamma^{\sigma}G_{0}(y-x)\rangle M^{\mu}_{\rho}M^{\nu}_{\sigma}A_{\nu}(y)-\nonumber\\
&-&i e^2\zeta^{0}\int d^{3}y\langle \gamma^{\rho} G_{0}(x-y)\gamma^{\sigma}\gamma_{S}G_{0}(y-x)\rangle M^{\mu}_{\rho}M^{\nu}_{\sigma}F_{0\nu}(y),
\end{eqnarray}
or, in momentum space:

\begin{eqnarray}
J^{\mu}(q)=-ie^{2}\Pi^{\rho\sigma}(\tilde{q})M^{\mu}_{\rho}M^{\nu}_{\sigma}A_{\nu}(q)-ie^{2}\zeta^{0}\Pi^{\rho\sigma}_{S}(\tilde{q})M^{\mu}_{\rho}M^{\nu}_{\sigma}F_{0\nu}(q),
\end{eqnarray}

where we have defined the following polarization functions:

\begin{subequations}
\begin{equation}
\Pi^{\rho\sigma}(\tilde{q})=\frac{1}{v^{2}}\int \frac{d^3 \tilde{k}}{(2\pi)^3}Tr \left[\gamma^{\rho}G_{0}(\tilde{k})\gamma^{\sigma}G_{0}(\tilde{k}-\tilde{q})\right],
\end{equation}
\begin{equation}
\Pi^{\rho\sigma}_{S}(\tilde{q})=\frac{1}{v^2}\int \frac{d^3 \tilde{k}}{(2\pi)^3}Tr \left[\gamma^{\rho}G_{0}(\tilde{k})\gamma^{\sigma}\gamma_{S}G_{0}(\tilde{k}-\tilde{q})\right],
\end{equation}
\end{subequations}
with $\tilde{k}^{\mu}\equiv M^{\mu}_{\nu}k^{\nu}=(k_{0},vk_{1},vk_{2})$. The fermionic Green function in momentum space takes the form 
\begin{equation}
G_{0}(\tilde{k})=\frac{\gamma^{\alpha}\tilde{k}_{\alpha}+m}{\tilde{k}^2+m^2}.
\end{equation}

In what follows we will omit the tilde in all the expressions, remembering all external momenta in the correlation functions should be replaced by the momenta with tilde at the end of the calculations. Applying the standard technology of the computation of Feynman diagrams, and knowing that ($\eta^{\rho\sigma}=$diag$(1,-1,-1)$ is the metric tensor in three space-time dimensions)
\begin{subequations}
\begin{equation}
Tr\left[\gamma^{\rho}\gamma^{\alpha}\gamma^{\sigma}\gamma^{\beta}\right]=4\left(\eta^{\rho\alpha}\eta^{\sigma\beta}-
\eta^{\rho\sigma}\eta^{\alpha\beta}+\eta^{\rho\beta}\eta^{\sigma\alpha}\right),
\end{equation}
\begin{equation}
Tr\left[\gamma^{\rho}\gamma^{\sigma}\right]=4\eta^{\rho\sigma},
\end{equation}
\begin{equation}
Tr\left[\gamma^{\rho}\gamma^{\alpha}\gamma_{S}\gamma^{\sigma}\right]=4i \varepsilon^{\rho\alpha\sigma},
\end{equation}
\end{subequations}
the polarization functions take the final form
\begin{subequations}
\begin{equation}
\Pi^{\rho\sigma}(\tilde{q})=\frac{i}{v^{2}\pi}\int^{1}_{0}dx \frac{x(x-1)}{\sqrt{m^2+x(x-1)\tilde{q}^{2}}}\left(\tilde{q}^{\rho}\tilde{q}^{\sigma}-\tilde{q}^{2}\eta^{\rho\sigma}\right),
\end{equation}
\begin{equation}
\Pi_{S}^{\rho\sigma}(\tilde{q})=\frac{m}{2\pi v^{2}}\int^{1}_{0}dx\frac{1}{\sqrt{m^2+x(x-1)\tilde{q}^{2}}}\varepsilon^{\rho\alpha\sigma}\tilde{q}_{\alpha}.
\end{equation}
\end{subequations}

The spatial components of the current density $J^{\mu}$, in the gauge $A_{0}=0$, $q_{0}\equiv\omega$, and after restoring $\hbar$
\begin{equation}
J^{i}(q)=\frac{e^2}{\pi\hbar}\left(v^{2}q^{i}q^{j}-v^{2}q^{2}\delta^{ij}+\omega^{2}\delta^{ij}\right)\Pi_{e}(\omega,\mathbf{q})A_{j}(q)-\frac{ie^2}{2\pi\hbar}m\zeta^{0}\Pi_{o}(\omega,\mathbf{q})\varepsilon^{ij}\omega F_{0j},
\end{equation}
with 

\begin{equation}
\Pi_{e}(\omega,\mathbf{q})=\int^{1}_{0}dx \frac{x(x-1)}{\sqrt{m^2+x(x-1)(\omega^{2}-v^2q^{2})}},
\end{equation}
and 
\begin{equation}
\Pi_{o}(\omega,\mathbf{q})=\int^{1}_{0}dx \frac{1}{\sqrt{m^2+x(x-1)(\omega^{2}-v^2q^{2})}}.
\end{equation}

It is important to note here that all the dependence with the spatial components of the momentum $q$ is quadratic. It implies that in the electromagnetic response of the system reflects the fact that there is no breakdown of inversion symmetry, ruling out the possibility that the optical activity described in the main text has its origin in the breakdown of such discrete symmetry.

Said that, we will be interested in the long wavelength limit of the electromagnetic response, performing the limit $q\rightarrow 0$. The current density simplifies to 

\begin{equation}
J^{i}(q)=\frac{e^2}{\pi\hbar}\omega^{2}\delta^{ij}\Pi_{e}(\omega)A_{j}(q)-\frac{ie^2}{2\pi\hbar}m\zeta^{0}\Pi_{o}(\omega)\omega\varepsilon^{ij} F_{0j},
\end{equation}
with (for simplicity we will set $m>0$)
\begin{equation}
\Pi_{e}(\omega)=\int^{1}_{0}dx \frac{x(x-1)}{\sqrt{m^2+x(x-1)\omega^{2}}}=\frac{m}{2\omega^2}-\frac{(4m^2+\omega^2)}{8\omega^{3}}\log\left\vert\frac{\omega+2m}{\omega-2m}\right\vert+i\pi\frac{(4m^2+\omega^2)}{8\omega^{3}}\Theta(\omega-2m),
\end{equation}
and 
\begin{equation}
\Pi_{o}(\omega)=\int^{1}_{0}dx \frac{1}{\sqrt{m^2+x(x-1)\omega^{2}}}=\frac{1}{\omega}\log\left\vert\frac{\omega+2m}{\omega-2m}\right\vert-i\pi\frac{1}{\omega}\Theta(\omega-2m).
\end{equation}

These expressions are well known expressions for the susceptibilities in the context of massive QED$_{2+1}$\cite{R841,R842}. As it is known, these susceptibilities develop an imaginary part for frequencies $\omega$ larger than the system's gap $2m$.

In the chosen gauge, $E_{j}=F_{0j}=-i\omega A_{j}$, or $A_{j}=i E_{j}/\omega$, the the current reads

\begin{equation}
J^{i}=\frac{\alpha}{\pi}(i\omega \Pi_{e}(\omega))\delta^{ij}E_{j}-\frac{\alpha}{\pi}\Pi_{o}(\omega)im\zeta^{0}\omega\varepsilon^{ij}E_{j},
\end{equation}
or
\begin{equation}
J^{i}=\frac{\alpha}{\pi}\sigma_{xx}(\omega)\delta^{ij}E_{j}-\frac{\alpha}{\pi}i\zeta^{0}\omega\sigma_{xy}(\omega)\varepsilon^{ij}E_{j},
\end{equation}
where we have defined the longitudinal and Hall conductivities
\begin{subequations}
\begin{equation}
\sigma_{xx}(\omega)=i\omega \Pi_{e}(\omega),\label{sigmaxx}
\end{equation}
\begin{equation}
\sigma_{xy}(\omega)=m\Pi_{o}(\omega).\label{sigmaxy}
\end{equation}
\end{subequations}
There is a comment to be done here. For massive Dirac fermions in the continuum, the conductivity $\sigma_{xy}$ is half of an integer in the limit $\omega\rightarrow 0$, while it is well known that the conductivity should be an integer. This is an artefact of the continuum description, and the other half of $\sigma_{xy}$ comes from the states of the rest of the Brillouin zone. We have cured this deficiency by hand just multiplying $\sigma_{xy}$ by a factor $2$.

\subsection{Computation of the transmitted and reflected electromagnetic waves}

Here we will closely follow the steps done in reference \cite{FV09}. We shall consider our two dimensional system placed at the plane $z=0$. The Maxwell equations are of the form, in Gaussian units, 
\begin{equation}
\partial^{\mu}F_{\mu\nu}=-\frac{4\pi}{c}J_{\nu}\delta(z),
\end{equation}

Away from the plane $z=0$, the Maxwell equations describe the propagation of free electromagnetic waves. Let us consider a monocromatic wave travelling from the left to the right with momentum $\mathbf{q}=(0,0,q)$ and frequency $\omega=c q$ and polarized along the direction $O\hat{X}$. In this field configuration, apart from the gauge choice $A_{0}=0$, due to fact that both the electric and magnetic fields are perpendicular to $\mathbf{q}$, it is found that $A_{z}=0$ as well. The electromagnetic field at the left ($z<0$) of the plane $z=0$ will consist on the superposition of the incident wave together with the reflected wave of momentum $-q$ and polarization $(r_{xx},r_{xy})$. At $z>0$ the transmitted wave will have momentum $q$ and polarization $(t_{xx},t_{xy})$.

To determine the coefficients $t_{xx}, t_{xy}, r_{xx}$, and $r_{xy}$ we have to apply the boundary conditions $A_{i}(z\rightarrow 0^{-})=A_{i}(z\rightarrow 0^{-})$ and 
\begin{equation}
\left[\frac{dA_{i}}{d z}\right]_{z\rightarrow 0^{+}}-\left[\frac{dA_{i}}{d z}\right]_{z\rightarrow 0^{-}}=\frac{4\pi}{c}\left[J_{i}(q)\right]_{z=0},
\end{equation}
which translates into the following algebraic system:
\begin{eqnarray}
\left( \begin{array}{cccc}
1 & 0 & -1 & 0 \\
0 & 1 & 0 & -1 \\
1+4\alpha\sigma_{xx} & -4\alpha i\zeta^{0}\omega\sigma_{xy} & 1 & 0 \\
4\alpha i\zeta^{0}\omega\sigma_{xy} & 1+4\alpha\sigma_{xx} & 0 & 1 \end{array} \right)\left(\begin{array}{c}
t_{xx}\\
t_{xy}\\
r_{xx}\\
r_{xy}\end{array}\right)=\left(\begin{array}{c}
1\\
0\\
1\\
0\end{array}\right).
\end{eqnarray}
The solution of this system is:
\begin{subequations}
\begin{equation}
t_{xx}=\frac{1+2\alpha \sigma_{xx}}{(1+2\alpha\sigma_{xx})^2-4\alpha^{2}\zeta^{02}\omega^{2}\sigma^{2}_{xy}},
\end{equation}
\begin{equation}
t_{xy}=-\frac{2i\alpha\zeta^{0}\omega\sigma_{xy}}{(1+2\alpha\sigma_{xx})^2-4\alpha^{2}\zeta^{02}\omega^{2}\sigma^{2}_{xy}},
\end{equation}
\begin{equation}
r_{xx}=-1+\frac{1+2\alpha \sigma_{xx}}{(1+2\alpha\sigma_{xx})^2-4\alpha^{2}\zeta^{02}\omega^{2}\sigma^{2}_{xy}},
\end{equation}
\begin{equation}
r_{xy}=-\frac{2i\alpha\zeta^{0}\omega\sigma_{xy}}{(1+2\alpha\sigma_{xx})^2-4\alpha^{2}\zeta^{02}\omega^{2}\sigma^{2}_{xy}}.
\end{equation}
\end{subequations}
An important simplification is possible if, for frequencies $\omega$ smaller than the gap $2m$, we perform an expansion in terms of $\alpha$. To lowest order in $\alpha$,
The previous coefficients take the simpler form 

\begin{subequations}
\begin{equation}
t_{xx}\simeq 1,
\end{equation}
\begin{equation}
t_{xy}\simeq -2i\alpha \zeta^{0}\omega Re[\sigma_{xy}],
\end{equation}
\begin{equation}
r_{xx}\simeq-2i\alpha Im[\sigma_{xx}],
\end{equation}
\begin{equation}
r_{xy}\simeq -2i\alpha \zeta^{0}\omega Re[\sigma_{xy}].
\end{equation}
\end{subequations}

because in this regime $\omega<2m$, $Re[\sigma_{xx}]=Im[\sigma_{xy}]=0$. Also, the previous expressions are calculated as if the surrounding medium were the vacuum. If the medium is a material medium with a dielectric constant $\epsilon$, it is enough to make the change $\alpha\rightarrow \alpha/\sqrt{\epsilon}$.

In general, for any frequency, the coefficients $t_{ij}$ and $r_{ij}$ will be complex functions of $\omega$ so the polarization of both the transmitted and reflected waves will be elliptical. The angle of polarization rotation can be calculated as

\begin{equation}
\theta_{r}=\frac{1}{2}\left(\arg(r_{xx}+i r_{xy})-\arg(r_{xx}-i r_{xy})\right)\equiv\frac{1}{2}\arg\left(\frac{r_{xx}+i r_{xy}}{r_{xx}-i r_{xy}}\right),
\end{equation}
and 
\begin{equation}
\theta_{t}=\frac{1}{2}\left(\arg(t_{xx}+i t_{xy})-\arg(t_{xx}-i t_{xy})\right)\equiv\frac{1}{2}\arg\left(\frac{t_{xx}+i t_{xy}}{t_{xx}-i t_{xy}}\right).
\end{equation}

The expressions of $\theta_{r}$ and $\theta_{t}$ in terms of $\sigma_{xx}$ and $\sigma_{xy}$ are

\begin{equation}
\theta_{r}=\frac{1}{2}\arg\left(\frac{(\sigma_{xx}-\zeta^{0}\omega\sigma_{xy})}{(\sigma_{xx}+\zeta^{0}\omega\sigma_{xy})}\frac{(2\alpha(\sigma_{xx}+\zeta^{0}\omega\sigma_{xy})-1)}{(2\alpha(\sigma_{xx}-\zeta^{0}\omega\sigma_{xy})-1)}\right),
\end{equation}
\begin{equation}
\theta_{t}=\frac{1}{2}\arg\left(\frac{2\alpha(\sigma_{xx}+\zeta^{0}\omega\sigma_{xy})-1}{2\alpha(\sigma_{xx}-\zeta^{0}\omega\sigma_{xy})-1}\right).
\end{equation}

\subsection{Demonstration of the non-reciprocity relation}

Let us stablish how the Lorentz reciprocity theorem gets modified due to the presence of the constitutive relations defined in the main text. Without loss of generality and in order to keep things as simple as it is possible, we will assume that there is no longitudinal response due to the QW ($\sigma_{xx}=0$) and the surrounding medium has the vacuum polarizability, so $\epsilon=1$. Under these circumstances, the displacement field and the magnetic field read
\begin{subequations}
\begin{equation}
4\pi D_{i}=E_{i}+\theta f(x) \dot{B}_{i},
\end{equation}
\begin{equation}
4\pi H_{i}=B_{i}-\theta f(x)\dot{E}_{i},
\end{equation}
\label{constrel}
\end{subequations}
whre $f(x)$ is the same function defined in the main text, and the parameter $\theta$ is $\theta=-\zeta^{0}\sigma_{xy}$, and $c=1$ to ease notation.

We will make extensive use of the two following Maxwell equations:
\begin{subequations}
\begin{equation}
\nabla\times\mathbf{E}=-\dot{\mathbf{B}},\label{ME1}
\end{equation}
\begin{equation}
\nabla\times\mathbf{H}=4\pi \mathbf{J}+\dot{\mathbf{D}}.\label{ME2}
\end{equation}
\end{subequations}
Now, let us consider two different external currents denoted by the labels $\mathbf{J}_{1}$ and $\mathbf{J}_{2}$ placed at different points of the space and having the same harmonic time dependence $e^{i\omega t}$. The corresponding electric and magnetic fields induced by these currents will be denoted with the same subscripts. 

Let us take the scalar product of $\mathbf{H}_{2}$ and the equation (\ref{ME1}) for the electromagnetic field $1$, the scalar product of $-\mathbf{E}_{1}$ and the equation (\ref{ME2}) for the field $2$, the scalar product of $\mathbf{E}_{2}$ and the equation (\ref{ME2}) for the field $1$, and the scalar product of $-\mathbf{H}_{1}$ and the equation (\ref{ME1}) for the field $2$. Summing all together, going to the frequency domain, and making use of the relations (\ref{constrel}) we have
\begin{eqnarray}
\nabla\cdot\left(\mathbf{E}_{1}\times\mathbf{H}_{2}-\mathbf{E}_{2}\times\mathbf{H}_{1}\right)
 =4\pi\left(\mathbf{E}_{2}\cdot\mathbf{J}_{1}-\mathbf{E}_{1}\cdot\mathbf{J}_{2}\right)+8\pi \omega^{2}\theta\left(\mathbf{E}_{1}\cdot\mathbf{H}_{2}-\mathbf{E}_{2}\cdot\mathbf{H}_{1}\right).
\end{eqnarray}
Integrating over all the space $\Omega$, we can make use of the Gauss theorem to transform the left hand side in a surface integral, leading to
\begin{eqnarray}
\int_{\partial\Omega} d^{2}S\hat{\mathbf{n}}\cdot\left(\mathbf{E}_{1}\times\mathbf{H}_{2}-\mathbf{E}_{2}\times\mathbf{H}_{1}\right)=
4\pi\int_{\Omega}d^{3}\mathbf{r}\left[\left(\mathbf{E}_{2}\cdot\mathbf{J}_{1}-\mathbf{E}_{1}\cdot\mathbf{J}_{2}\right)+2\omega^{2}\zeta^{0}\sigma_{xy} f(x)\left(\mathbf{E}_{1}\cdot\mathbf{H}_{2}-\mathbf{E}_{2}\cdot\mathbf{H}_{1}\right)\right].
\end{eqnarray}
It can be shown that the left hand side of the previous identity can be dropped, leading to the desired result, that is, the system described here does not satisfy the standard reciprocity theorem\cite{LLP60}.

For completeness, we quote the equivalent non reciprocity result that occurs in the Quantum Anomalous Hall effect, or more generally, to a generic magnetoelectric medium\cite{R73}:
\begin{eqnarray}
\int_{\partial\Omega} d^{2}S\hat{\mathbf{n}}\cdot\left(\mathbf{E}_{1}\times\mathbf{H}_{2}-\mathbf{E}_{2}\times\mathbf{H}_{1}\right)=
4\pi\int_{\Omega}d^{3}\mathbf{r}\left[\left(\mathbf{E}_{2}\cdot\mathbf{J}_{1}-\mathbf{E}_{1}\cdot\mathbf{J}_{2}\right)-2i\omega\sigma_{xy} f(x)\left(\mathbf{E}_{1}\cdot\mathbf{H}_{2}-\mathbf{E}_{2}\cdot\mathbf{H}_{1}\right)\right].
\end{eqnarray}


\end{widetext}

\end{document}